\documentclass[aps,pra,preprint,groupedaddress,superscriptaddress]{revtex4-2} 
\usepackage[colorlinks,linkcolor=red,anchorcolor=blue,citecolor=red]{hyperref} 
\usepackage{braket} 
\usepackage{amsmath} 
\usepackage{amssymb} 
\usepackage[dvips]{graphicx}
\usepackage{color}
\usepackage{verbatim}
\usepackage[T1]{fontenc}
\usepackage{siunitx} 
\usepackage{array}
\usepackage{multirow} 
\usepackage{booktabs} 

\newcommand{\ii}{\mathrm{i}}
\newcommand{\e}{\mathrm{e}}
\newcommand{\dd}{\mathrm{d}}
\makeatletter

\newcommand{\Rmnum}[1]{\expandafter\@slowromancap\romannumeral #1@}
\makeatother 

\DeclareMathOperator{\sinc}{sinc}

\begin{document}


\title{Precise Phase Error Rate Analysis for Quantum Key Distribution with Phase Postselection}



\author{Yao Zhou}
\affiliation{CAS Key Laboratory of Quantum Information, University of Science and Technology of China, Hefei, Anhui 230026, China}
\affiliation{CAS Center for Excellence in Quantum Information and Quantum Physics, University of Science and Technology of China, Hefei, Anhui 230026, China}

\author{Zhen-Qiang Yin}
\email{yinzq@ustc.edu.cn}
\affiliation{CAS Key Laboratory of Quantum Information, University of Science and Technology of China, Hefei, Anhui 230026, China}
\affiliation{CAS Center for Excellence in Quantum Information and Quantum Physics, University of Science and Technology of China, Hefei, Anhui 230026, China}
\affiliation{Hefei National Laboratory, University of Science and Technology of China, Hefei 230088, China}

\author{Yang-Guang Shan}
\author{Ze-Hao Wang}
\affiliation{CAS Key Laboratory of Quantum Information, University of Science and Technology of China, Hefei, Anhui 230026, China}
\affiliation{CAS Center for Excellence in Quantum Information and Quantum Physics, University of Science and Technology of China, Hefei, Anhui 230026, China}

\author{Shuang Wang}
\email{wshuang@ustc.edu.cn}
\author{Wei Chen}
\author{Guang-Can Guo}
\author{Zheng-Fu Han}
\affiliation{CAS Key Laboratory of Quantum Information, University of Science and Technology of China, Hefei, Anhui 230026, China}
\affiliation{CAS Center for Excellence in Quantum Information and Quantum Physics, University of Science and Technology of China, Hefei, Anhui 230026, China}
\affiliation{Hefei National Laboratory, University of Science and Technology of China, Hefei 230088, China}




\begin{abstract}
Quantum key distribution (QKD) stands as a pioneering method for establishing information-theoretically secure communication channels by utilizing the principles of quantum mechanics. In the security proof of QKD, the phase error rate serves as a critical indicator of information leakage and directly influences the security of the shared key bits between communicating parties, Alice and Bob. In estimating the upper bound of the phase error rate, phase randomization and subsequent postselection mechanisms serve pivotal roles across numerous QKD protocols.
Here we make a precise phase error rate analysis for QKD protocols with phase postselection, which helps us to accurately bound the amount of information an eavesdropper may obtain. We further apply our analysis in sending-or-not-sending twin-field quantum key distribution (SNS-TFQKD) and mode-pairing quantum key distribution (MP-QKD). The simulation results confirm that our precise phase error analysis can noticeably improve the key rate performance especially over long distances in practice. Note that our method does not require alterations to the existing experimental hardware or protocol steps. It can be readily applied within current SNS-TF-QKD and MP-QKD for higher key rate generation.
\end{abstract}


\maketitle

\section{Introduction}
Quantum key distribution (QKD) \cite{BB84,PhysRevLett.67.661} provides the unconditional secure keys, which can not be break even if the eavesdropper Eve has unlimited computing resources, between two remote parties by exploiting the fundamental properties of quantum physics. During the past four decades,
QKD has achieved great development in terms of security \cite{RN155,doi:10.1126/science.283.5410.2050,PhysRevLett.85.441,RennerPhD,RN127,PhysRevLett.98.140502,RN103,RN124,RN87,PhysRevResearch.3.023019,RevModPhys.94.025008} and practicality \cite{PhysRevLett.91.057901,PhysRevLett.94.230503,PhysRevLett.94.230504,PhysRevLett.108.130503,PhysRevLett.108.130502,RN167,RN74,PhysRevX.8.031043,PhysRevA.98.062323,RN128,RN90,RN129,RN166,PhysRevApplied.18.054026,PhysRevA.101.042330,RN81,PhysRevA.107.032621,RN178,PRXQuantum.3.020315,RN330}. 
The decoy-state method \cite{PhysRevLett.91.057901,PhysRevLett.94.230503,PhysRevLett.94.230504} allows QKD systems to utilize coherent optical sources, diverging from the standard single-photon source BB84 protocol. This adaptation renders practical QKD systems resilient against photon number splitting (PNS) attacks, significantly enhancing both the secure key rate and the achievable communication distance. 
Measurement-device-independent (MDI) QKD protocol \cite{PhysRevLett.108.130503} (see also \cite{PhysRevLett.108.130502}) designates the measurement party as an untrusted intermediary situated within the channel, thereby making the key bits shared between two communication parties immune to all detector side attacks.
However, due to the inherent transmission loss in the channel, the key rate performance in previous QKD is naturally constrained by the PLOB rate-transmittance bound \cite{RN88} (see also the TGW bound \cite{RN115}). The pursuit of longer communication distance and higher secure key rate is the central issue of practical QKD. Based on the simple and promising MDI-QKD structure, twin-field quantum key distribution (TF-QKD) \cite{RN74} and mode-pairing quantum key distribution (MP-QKD) \cite{RN178} (also named asynchronous-MDI-QKD \cite{PRXQuantum.3.020315}) were proposed to break the PLOB bound without quantum repeaters in recent years. Currently, these special variants substantially extend the point-to-point transmission distance, significantly advancing the practicality of QKD for longer-distance applications.

Roughly speaking, most QKD protocols consist of code mode and decoy
mode. The communicating parties Alice and Bob generate the raw keys in the code mode and disclose a part of raw keys to estimate the bit error rate for error correction step. The key information leakage of QKD can be bounded by the so-called phase error rate \cite{doi:10.1126/science.283.5410.2050,PhysRevLett.85.441,RN127}, which can be estimated in the decoy mode. From the perspective of the equivalent entanglement-based scheme for the actual QKD protocol, the key is obtained by measuring the bipartite auxiliary qubits AB in the $Z=\{\ket{0}, \ket{1}\}$ basis. The phase error rate is usually defined as the bit error rate of qubits AB in the $X=\{\ket{+}=\frac{1}{\sqrt{2}}(\ket{0}+\ket{1}), \ket{-}=\frac{1}{\sqrt{2}}(\ket{0}-\ket{1})\}$ basis. Nevertheless, we find that in certain QKD protocols with phase postselection, the definition of phase error rate occurs not in the X basis but rather in the conjugate basis $X_\delta=\{\ket{+\delta}=\frac{1}{\sqrt{2}}(\ket{0}+\e^{-\ii\delta}\ket{1}), \ket{-\delta}=\frac{1}{\sqrt{2}}(\ket{0}-\e^{-\ii\delta}\ket{1})\}$ $(\delta\in[0,2\pi))$. The distinct definitions of phase error rate across various conjugate bases commonly yield differing values, which prompts us to make a precise phase error rate analysis.

Based on the above idea, we propose a precise phase error rate analysis in this paper to further reduce the lower bound of phase error rate. Our method demonstrates noticeable enhancements in key rate performance for several QKD protocols with phase postselection, such as sending-or-not-sending twin-field quantum key distribution (SNS-TFQKD) \cite{PhysRevA.98.062323} and MP-QKD.

We structure the remainder of this paper as follows. In Sec.\ref{eph analysis}, we introduce a general equivalent entanglement-based scheme applicable to certain MDI-QKD variants and perform the precise phase error rate analysis. In Sec.\ref{apply}, we give the security proof for SNS-TF-QKD based on the equivalent entanglement-based scheme and obtain a precise phase error rate from the previously established formula. Our simulations demonstrate the noticeable enhancement achieved by our method in practical AOPP-SNS-TFQKD protocol. Additionally, we provide a brief overview of MP-QKD and showcase the improvements facilitated by our approach. Finally, a conclusion is given and we expect our method can be used in current AOPP-SNS-TFQKD and MP-QKD protocols.

\section{The precise phase error analysis}\label{eph analysis}
\begin{figure}
	\centering
	\includegraphics[width=\textwidth]{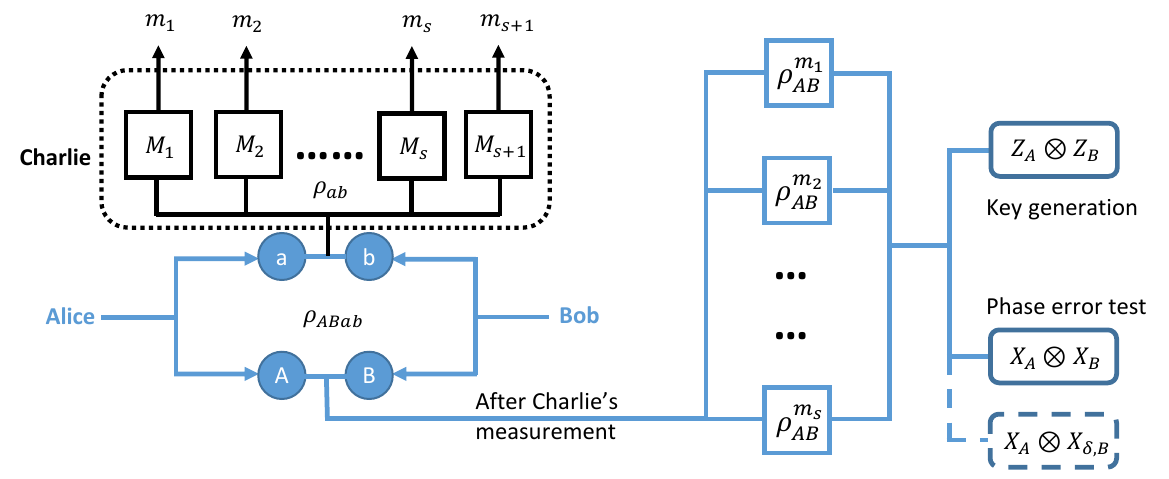}
	\caption{The equivalent entanglement-based scheme of certain MDI-QKD variants. Alice and Bob first prepare the ancillary qubits A, B and signal states a, b. The signal states a, b are sent to untrusted node Charlie for measurement. After Charlie announces the measurement outputs $m_i$ ($i=1,2,...,s,s+1$ and $s\ge 2$), Alice and Bob choose the measurement outcome $m_j$ ($j=1,2,...,s$) to generate the key. Note that the measurement outcome $m_{s+1}$ denotes an invalid event, which is not used to generate keys. In the key generation step, Alice and Bob measure the ancillary bipartite qubits $\rho_\text{AB}^{m_j}$ ($j=1,2,...,s$) in the $Z_\text{A}$ and $Z_\text{B}$ basis. The phase error rate is usually defined as the bit error rate of qubits $\rho_\text{AB}^{m_j}$ in the $X_\text{A}$ and $X_\text{B}$ ($X=\{\ket{+}=\frac{1}{\sqrt{2}}(\ket{0}+\ket{1}), \ket{-}=\frac{1}{\sqrt{2}}(\ket{0}-\ket{1})\}$) basis. In fact, it can also be defined in the $X_\text{A}$ and $X_{\delta,\text{B}}$ ($X_\delta=\{\ket{+\delta}=\frac{1}{\sqrt{2}}(\ket{0}+\e^{-\ii\delta}\ket{1}), \ket{-\delta}=\frac{1}{\sqrt{2}}(\ket{0}-\e^{-\ii\delta}\ket{1})\}$) basis.
	}
	\label{MDI}
\end{figure}

We first consider the following equivalent entanglement-based scheme for certain MDI-QKD variants in fig.\ref{MDI}.
The communicating parties Alice and Bob prepare the ancillary qubit particles $A$, $B$ and the signal particles $a$, $b$ in a joint quantum state $\rho_\text{ABab}$. The signal particles $a$ and $b$ are sent to untrusted party Charlie through an untrusted channel. Charlie measures the received signal states and announces the measurement outputs $m_i$ ($i=1,2,...,s,s+1$ and $s\ge 2$). Without loss of generality, we can introduce a positive operator-valued measure (POVM), which is a set of positive semi-definite Hermitian matrices $\{M_1, M_2, ..., M_s, M_{s+1}\}$ acting on state $\rho_\text{ab}=\text{Tr}_\text{AB}(\rho_\text{ABab})$ associated with the outcomes $\{m_1, m_2,..., m_s, m_{s+1}\}$, to denote the Charlie's measurement and channel transmission effects. So the probability $p(m_i)$ of the outcome $m_i$ is $p(m_i)=\text{Tr}(\rho_\text{ab}M_i^\dagger M_i)$. After Charlie's measurement and announcing the outcome $m_i$, the ancillary bipartite qubits collapse into the normalized quantum state $\rho_\text{AB}^{m_i}=\text{Tr}_\text{ab}(\rho_\text{ABab}M_i^\dagger M_i))/p(m_i)$. Note that the measurement outcome $m_{s+1}$ denotes an invalid event, which is not used to generate keys. For the ancillary bipartite qubits $\rho_\text{AB}^{m_j}$ ($j=1,2,...,s$), Alice and Bob both measure them in the $Z=\{\ket{0},\ket{1}\}$ basis to obtain the key bits or measure them in the $X=\{\ket{+}=\frac{\ket{0}+\ket{1}}{\sqrt{2}},\ket{-}=\frac{\ket{0}-\ket{1}}{\sqrt{2}}\}$ basis for phase error test. If Alice still measures her ancillary qubit in the $X$ basis but Bob measures his ancillary qubit in the $X_\delta=\{\ket{+\delta}=\frac{1}{\sqrt{2}}(\ket{0}+\e^{-\ii\delta}\ket{1}), \ket{-\delta}=\frac{1}{\sqrt{2}}(\ket{0}-\e^{-\ii\delta}\ket{1})\}$ basis, 
it can still be used for phase error test. In fact, we can introduce a unitary operator $U_\text{B}^\delta$ for Bob's qubit B to convert the ancillary bipartite qubits $\rho_\text{AB}^{m_j}$ to $\sigma_\text{AB}^{m_j}$ where $U_\text{B}^\delta\ket{0}_\text{B}=\ket{0}_\text{B}$ and $U_\text{B}^\delta\ket{1}_\text{B}=\e^{\ii \delta}\ket{1}_\text{B}$. As discussed in ref.\cite{RN330}, this unitary operator has no physical effects on the key bits generation and Eve's potential system. It is obvious that the measurement output for $\sigma_\text{AB}^{m_j}$ in the $X$ and $X$ basis can be defined as the phase error. So we conclude that the phase error rate can be obtained by measuring the $\rho_\text{AB}^{m_j}$ in the $X$ and $X_\delta$ basis due to the fact that such measurement output is equivalent to measuring $\sigma_\text{AB}^{m_j}$ in the $X$ and $X$ basis.

In fact, we can separately define the phase error rates of the key bits under every effective announcement by Charlie. That means we can classify the key bits into $s$ classes according to the announced outputs by Charlie. We denote the phase error rate of $\rho_\text{AB}^{m_j}$ under the $X$ and $X$ basis as $e_\text{ph}^j$ and under the $X$ and $X_\delta$ basis as $e_\text{ph}^{\delta,j}$. We will estimate the upper bound on $e_\text{ph}^j$ given $e_\text{ph}^{\delta,j}$ below.

For the given ancillary bipartite qubits $\rho_\text{AB}^{m_{j}}$, Alice first measures her local qubit in the $X$ basis and obtain the state $\ket{+}$ with probability $p_+$ and $\ket{-}$ with probability $p_-$ ($p_+ + p_-=1$). Bob also measures his local qubit in the $X$ basis. Given Alice's output $\ket{+}$, we can assume that Bob obtains the state $\ket{+}$ with probability $1-e_+$ and $\ket{-}$ with probability $e_+$. Given Alice's output $\ket{-}$, we assume that Bob obtains the state $\ket{-}$ with probability $1-e_-$ and $\ket{+}$ with probability $e_-$ . Here, we have the phase error rate 
\begin{align}
	e_\text{ph}^j=p_+ e_+ + p_- e_-.
\end{align}
Given Alice's output $\ket{+}$, Bob's density matrix is
\begin{align}
	\rho_\text{B}^+=(1-e_+)\ket{+}\bra{+}+e_+\ket{-}\bra{-}+x_+\ket{+}\bra{-}+x_+^\star\ket{-}\bra{+},
\end{align}
where $x_+$ is a complex number, $x_+^\star$ is the complex conjugate of $x_+$. If Bob measures $\rho_\text{B}^+$ in the $X_\delta$ basis and defines the output $\ket{-\delta}$ as the error event, the error rate is
\begin{align}
	e_+^\delta
	&=\bra{-\delta}\rho_\text{B}^+\ket{-\delta}\nonumber\\
	&=(1-e_+)\frac{1-\cos\delta}{2}+e_+\frac{1+\cos\delta}{2}+x_+\frac{1+\e^{-\ii\delta}}{2}\frac{1-\e^{\ii\delta}}{2}+x_+^\star\frac{1-\e^{-\ii\delta}}{2}\frac{1+\e^{\ii\delta}}{2}\nonumber\\
	&=e_+\cos\delta+\frac{1-\cos\delta}{2}+\mathrm{Re}[-\ii x_+ \sin\delta],
\end{align}
where $\mathrm{Re}[x]$ is the real part of the complex number $x$. Given Alice's output $\ket{-}$, Bob's density matrix is
\begin{align}
	\rho_\text{B}^-=e_-\ket{+}\bra{+}+(1-e_-)\ket{-}\bra{-}+x_-\ket{+}\bra{-}+x_-^\star\ket{-}\bra{+},
\end{align}
where $x_-$ is a complex number, $x_-^\star$ is the complex conjugate of $x_-$. If Bob measures $\rho_\text{B}^-$ in the $X_\delta$ basis and defines the output $\ket{+\delta}$ as the error event, the error rate is
\begin{align}
	e_-^\delta
	&=\bra{+\delta}\rho_\text{B}^-\ket{+\delta}\nonumber\\
	&=e_-\frac{1+\cos\delta}{2}+(1-e_-)\frac{1-\cos\delta}{2}+x_-\frac{1+\e^{\ii\delta}}{2}\frac{1-\e^{-\ii\delta}}{2}+x_-^\star\frac{1-\e^{\ii\delta}}{2}\frac{1+\e^{-\ii\delta}}{2}\nonumber\\
	&=e_-\cos\delta+\frac{1-\cos\delta}{2}+\mathrm{Re}[\ii x_- \sin\delta].
\end{align}
So the phase error rate $e_\text{ph}^{\delta,j}$ under the $X$ and $X_\delta$ basis is
\begin{align}
	e_\text{ph}^{\delta,j}
	&=p_+e_+^\delta+p_-e_-^\delta \nonumber\\
	&=e_\text{ph}^j\cos\delta+\frac{1-\cos\delta}{2}+p_+\mathrm{Re}[-\ii x_+ \sin\delta]+p_-\mathrm{Re}[\ii x_- \sin\delta] \nonumber\\
	&=e_\text{ph}^j\cos\delta+\frac{1-\cos\delta}{2}+A_X^j \sin\delta,
\end{align}
where $A_X^j=p_+\mathrm{Re}[-\ii x_+]+p_-\mathrm{Re}[\ii x_-]$. Note that $A_X^j$ is independent of $\delta$.

In most practical MDI-QKD variants, we do not consider the phase error rates under different Charlie's announcements but define only one phase error rate for all key bits. We assume that the probability of effective events $m_{j}$ announced by Charlie in an effective round is $p_{j}$ ($\sum_{j=0}^{s}p_{j}=1$). So the total phase error rate under the $X$ and $X$ basis is
\begin{align}
	e_\text{ph}=\sum_{j=0}^{s} p_{j} e_\text{ph}^j,
\end{align}
and the total phase error rate under the $X$ and $X_\delta$ basis is
\begin{align}
	e_\text{ph}^\delta
	&=\sum_{j=0}^{s} p_{j} e_\text{ph}^{\delta,j}\nonumber\\
	&=\sum_{j=0}^{s} p_{j} \bigg( e_\text{ph}^j\cos\delta+\frac{1-\cos\delta}{2}+A_X^j \sin\delta
	\bigg)\nonumber\\
	&=e_\text{ph}\cos\delta+\frac{1-\cos\delta}{2}+ (\sum_{j=0}^{s} p_{j}A_X^j)\sin\delta . \label{e_delta}
\end{align}
We find that the phase error rate in some certain QKD variants with phase postselection is defined as $e_\text{ph}^\Delta=\frac{1}{2\Delta}\int_{-\Delta}^{\Delta} e_\text{ph}^\delta \dd\delta$ ($0<\Delta<\frac{\pi}{2}$), which is easily estimated by the decoy-state analysis.
In fact, $e_\text{ph}$ and  $e_\text{ph}^\Delta$ have the following correlation
\begin{align}
	e_\text{ph}^\Delta
	&=\frac{1}{2\Delta}\int_{-\Delta}^{\Delta} e_\text{ph}^\delta \dd\delta \nonumber\\
	&= \frac{1}{2\Delta}\int_{-\Delta}^{\Delta} \bigg(e_\text{ph}\cos\delta+\frac{1-\cos\delta}{2}+ (\sum_{j=0}^{s} p_{j}A_X^j)\sin\delta\bigg) \dd\delta \nonumber\\
	&=\frac{1-\sinc \Delta}{2}+e_\text{ph}\sinc\Delta, \label{e_Delta}
\end{align}
where $\sinc(x)=\frac{\sin(x)}{x}$. So, we can estimate the precise phase error rate by the previous given phase error rate $e_\text{ph}^\Delta$
\begin{align}
	e_\text{ph}
	&= \frac{1}{\sinc\Delta}e_\text{ph}^{\Delta}+\frac{1}{2}\big(1-\frac{1}{\sinc\Delta}\big) \nonumber\\
	&\le \frac{1}{\sinc\Delta}\bar{e}_\text{ph}^{\Delta}+\frac{1}{2}\big(1-\frac{1}{\sinc\Delta}\big),
\end{align}
where $\bar{e}_\text{ph}^{\Delta}$ is the upper bound of previous loose phase error rate estimated by the decoy-state analysis. 
Note that our analysis is applicable to the finite-key regime as long as the previous loose phase error rate is also for the finite-key case.

\section{Some QKD with phase postselection applicable to our method} \label{apply}
We find some QKD protocols with phase postselection applicable to our precise phase error rate analysis. We aim to implement our method within SNS-TFQKD and MP-QKD protocols, simulating its potential enhancements.

\subsection{apply our method to SNS-TFQKD}
The SNS-TFQKD protocol, introduced by Wang et al. in 2018, has emerged as a prominent TF-QKD protocol in current practice. In code mode, Alice (Bob) generates a key bit 1 (0) with a probability of $p$ while sending a phase-randomized coherent state. Conversely, she (he) generates a key bit 0 (1) with a probability of $1-p$ and does not send anything.
We give the equivalent entanglement-based scheme as follows:
\begin{align}
	\rho_\text{ABab}&=\sqrt{1-p}\ket{0}_\text{A}\ket{0}_\text{a}+\sqrt{p}\ket{1}_\text{A}\ket{\sqrt{\mu}\e^{\ii\alpha}}_\text{a}
	\otimes
	\sqrt{p}\ket{0}_\text{B}\ket{\sqrt{\mu}\e^{\ii\beta}}_\text{b}+\sqrt{1-p}\ket{1}_\text{B}\ket{0}_\text{b} \nonumber\\
	&=\sqrt{p(1-p)}\ket{00}_\text{AB}\ket{0}_\text{a}\sum_{m=0}^{\infty}\sqrt{P_m}\e^{\ii m\beta}\ket{m}_\text{b} + \sqrt{p(1-p)} \ket{11}_\text{AB}\sum_{n=0}^{\infty}\sqrt{P_n}\e^{\ii n\alpha}\ket{n}_\text{a}\ket{0}_\text{b} \nonumber\\
	&\quad+ (1-p)\ket{01}_\text{AB}\ket{00}_\text{ab} +p\ket{10}_\text{AB}\sum_{n=0}^{\infty}\sqrt{P_n}\e^{\ii n\alpha}\ket{n}_\text{a}\sum_{m=0}^{\infty}\sqrt{P_m}\e^{\ii m\beta}\ket{m}_\text{b}
\end{align}
where $\ket{0}_\text{A(B)}$ and $\ket{1}_\text{A(B)}$ denote the local auxiliary qubits which are used to generate the key between Alice and Bob, $\ket{0}_\text{a(b)}$ denotes the vacuum state, $\ket{\sqrt{\mu}\e^{\ii\alpha(\beta)}}_\text{a(b)}$ denotes the phase-randomized coherent state sent by Alice (Bob), $P_n=\e^{-\mu}\mu^n/n!$ is the Poisson distribution with mean photon number $\mu$, and $\alpha$ ($\beta$) is the random phase.

In a round of code mode in SNS-TFQKD, the key bit shared between Alice and Bob when one side sends nothing and the other side happens to send out the single-photon pulse is defined as the untagged bit \cite{PhysRevA.98.062323}. 
Only the untagged bits are deemed as genuinely valid coded bits, originating from the partial quantum state $\rho_\text{ABab}$ within the corresponding entanglement-based scheme.
\begin{align}
	\rho_\text{ABab}^\text{u}&=\frac{\sqrt{p(1-p)}\ket{00}_\text{AB}\ket{0}_\text{a}\sqrt{P_1}\e^{\ii \beta}\ket{1}_\text{b} + \sqrt{p(1-p)} \ket{11}_\text{AB}\sqrt{P_1}\e^{\ii \alpha}\ket{1}_\text{a}\ket{0}_\text{b}}{\sqrt{2p(1-p)P_1}} \nonumber\\
	&=\frac{\ket{00}_\text{AB}\ket{0}_\text{a}\e^{\ii (\beta-\alpha)}\ket{1}_\text{b} + \ket{11}_\text{AB}\ket{1}_\text{a}\ket{0}_\text{b}}{\sqrt{2}},
\end{align}
where $\ket{1}_\text{a(b)}$ is the single-photon state when Alice (Bob) sends the coherent state. In fact, the relative phase between $\ket{00}_\text{AB}\ket{01}_\text{ab}$ and $\ket{11}_\text{AB}\ket{10}_\text{ab}$ plays no roles in the results of  the measurement for generating the secure key and Eve's potential system. As the method proposed in ref.\cite{RN330}, we can introduce a unitary operation $U_\text{AB}^{\alpha\beta}$ to the bipartite auxiliary qubits AB before the measurement on them, where $U_\text{AB}^{\alpha\beta}\ket{00}_\text{AB}=\e^{\ii (\alpha-\beta)}\ket{00}_\text{AB}$ and $U_\text{AB}^{\alpha\beta}\ket{11}_\text{AB}=\ket{11}_\text{AB}$. This unitary operation can be achieved by constructing a hypothetical private channel through which Alice and Bob can share phase information $\alpha$ and $\beta$. So the quantum state $\sigma_\text{ABab}^\text{u}$ can take the following equivalent form
\begin{align}
	\sigma_\text{ABab}^\text{u}=\frac{\ket{00}_\text{AB}\ket{01}_\text{ab} + \ket{11}_\text{AB}\ket{10}_\text{ab}}{\sqrt{2}}. \label{sigma^u_ABab}
\end{align}
We can reformulate $\sigma^\text{u}_\text{ABab}$ in the $X$ and $X_\delta$ basis as
\begin{align}
	\sigma^\text{u}_\text{ABab}
	&=\frac{\frac{\ket{+}_\text{A}+\ket{-}_\text{A}}{\sqrt{2}}\frac{\ket{+\delta}_\text{B}+\ket{-\delta}_\text{B}}{\sqrt{2}}\ket{01}_\text{ab}+\frac{\ket{+}_\text{A}-\ket{-}_\text{A}}{\sqrt{2}}\frac{\ket{+\delta}_\text{B}-\ket{-\delta}_\text{B}}{\sqrt{2}\e^{-\ii\delta}}\ket{10}_\text{ab}}{\sqrt{2}} \nonumber\\
	&=\frac{\frac{\ket{+}_\text{A}\ket{+\delta}_\text{B}+\ket{-}_\text{A}\ket{-\delta}_\text{B}}{\sqrt{2}}
		\frac{\ket{01}_\text{ab}+\e^{\ii\delta}\ket{10}_\text{ab}}{\sqrt{2}}+
		\frac{\ket{+}_\text{A}\ket{-\delta}_\text{B}+\ket{-}_\text{A}\ket{+\delta}_\text{B}}{\sqrt{2}}
		\frac{\ket{01}_\text{ab}-\e^{\ii\delta}\ket{10}_\text{ab}}{\sqrt{2}}}{\sqrt{2}}.
\end{align}
This indicates that the phase error rate defined in the $X$ and $X_\delta$ basis is related to the yields of the quantum states $\frac{\ket{01}_\text{ab}+\e^{\ii\delta}\ket{10}_\text{ab}}{\sqrt{2}}$ and $\frac{\ket{01}_\text{ab}-\e^{\ii\delta}\ket{10}_\text{ab}}{\sqrt{2}}$. 

In the decoy mode of SNS-TFQKD, Alice and Bob prepare and send the phase-randomized coherent state with intensity $\mu_1$ to Charlie. After Charlie's measurement and announcement, Alice and Bob disclose the phases $\theta_\text{A}$ and $\theta_\text{B}$ of each pulse and post-select the instances that $|\theta_A-\theta_B|\le\frac{\Delta}{2}$ and $|\theta_A-\theta_B-\pi|\le\frac{\Delta}{2}$ for phase error rate estimation \cite{RN75}. When the phase difference between Alice and Bob is $|\theta_A - \theta_B| = \delta$ or $|\theta_A - \theta_B - \pi| = \delta$, they can estimate the phase error rate $e_\text{ph}^\delta$ as depicted in (\ref{e_delta}).
So our precise phase error analysis is adapted to SNS-TFQKD protocol. In the decoy mode, ref.\cite{RN75} uses the round that Alice and Bob both send the coherent state with intensity $\mu_1$ to estimate the phase  error rate in (\ref{e_Delta}) as
\begin{equation}
	e_\text{ph}^\Delta\le \bar{e}_1^\text{ph}=\frac{T_\Delta-\frac{1}{2}\e^{-2\mu_1}S_{00}}{2\mu_1\e^{-2\mu_1}\underline{s}_1^\text{Z}}
\end{equation}
where $T_\Delta$ is the error click ratio of the instances that Alice and Bob both send the coherent pulse with intensity $\mu_1$ and their phase difference meets the post-selection condition: $|\theta_A-\theta_B|\le\frac{\Delta}{2}$ or $|\theta_A-\theta_B-\pi|\le\frac{\Delta}{2}$ \cite{RN75}, $S_{00}$ is the counting rate of vacuum sate and $\underline{s}_1^\text{Z}$ is the lower bound of the counting rate of single-photon state. So the precise phase error rate is
\begin{align}
	e_\text{ph}^\text{p}
	\le\frac{1}{\sinc(\frac{\Delta}{2})}\frac{T_\Delta-\frac{1}{2}\e^{-2\mu_1}S_{00}}{2\mu_1\e^{-2\mu_1}\underline{s}_1^\text{Z}}
	+\frac{1}{2}\big(1-\frac{1}{\sinc(\frac{\Delta}{2})}\big).
\end{align}

\begin{figure}
	\centering
	\includegraphics[width=\textwidth]{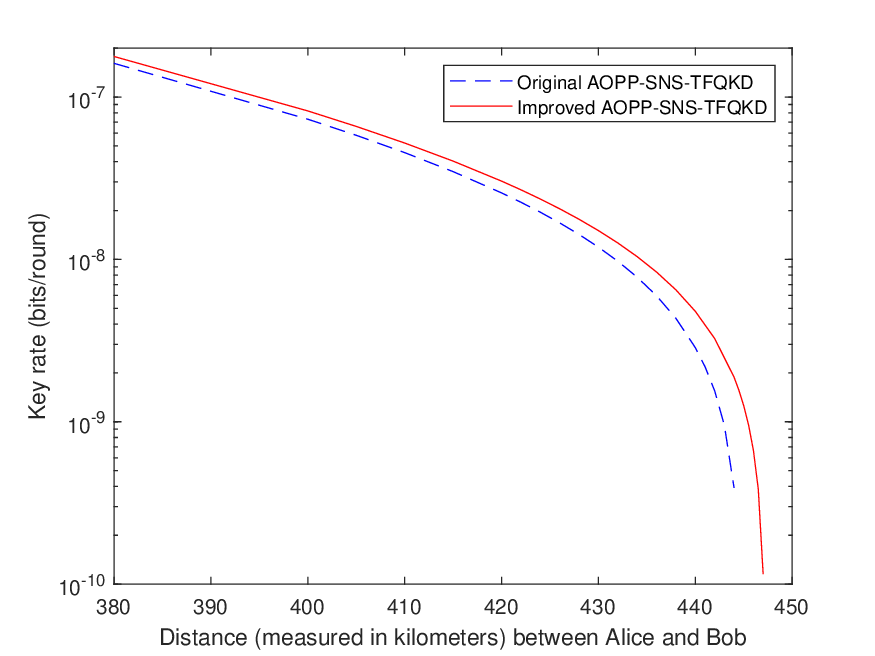}
	\caption{
		The key rate per round in AOPP-SNS-TFQKD, comparing the original protocol with the outcomes of our precise analysis.
	}
	\label{fig.AOPP-SNS-TFQKD}
\end{figure}

We use this precise phase error analysis in the practical AOPP-SNS-TFQKD protocol \cite{PhysRevA.101.042330,Jiang_2020,RN81} and simulate the original AOPP-SNS-TFQKD and our improved protocol in fig.\ref{fig.AOPP-SNS-TFQKD}.

We use the same finite-key analysis and linear simulation model mentioned in ref.\cite{RN81}. We set the total sending pulse as \num{1e12}, the misalignment error as 5\%, the dark count rate as \num{1e-8}, the
fiber loss coefficient as \SI{0.2}{dB/km}, the photon detection efficiency as 30\%, the error correction inefficiency as \num{1.1} and the failure probability when calculating the effect of statistical fluctuation as \num{1e-20}. Furthermore, we achieve a security level of \num{4.66e-9} in the sense of composable security against coherent attacks. 
The simulation results indicate a key rate improvement of over 10\% for distances beyond \SI{380}{km}, accompanied by an extended maximum achievable distance of \SI{3}{km}. These findings affirm the practical effectiveness of our precise phase error analysis in noticeably enhancing the key rate performance.

\subsection{apply our method to MP-QKD}
The specific process of MP-QKD protocol and the security proof in the finite-key regime based on the equivalent entanglement-based scheme have been thoroughly discussed in ref.\cite{RN330}. Here, we provide a brief overview of the coded quantum states sent by Alice and Bob and directly show the equivalent entanglement-based scheme.

In each round of MP-QKD, Alice (Bob) randomly sends the phase-randomized coherent pulses with intensity $\mu_{a(b)}$ and $\nu_{a(b)}$ and the vacuum state with probabilities $p_{\mu_{a(b)}}$, $p_{\nu_{a(b)}}$ and $p_o$ to untrusted node Charlie for interference measurement. Only the rounds where only detector L or R clicks are kept for subsequent step. In the post-processing step, Alice and Bob choose two rounds in the maximal pairing interval to form the effective event pair. We denote the intensity in the first and second round of the effective event pair as $k^1_{a(b)}$ and $k^2_{a(b)}$. Only those pairs in which vacuum states are paired with weak coherent states are used for key generation. Alice sets her key bit to 0 if $k^2_{a}\neq k^1_{a}=0$ or 1 if $k^1_{a}\neq k^2_{a}=0$. Bob sets her key bit to 0 if $k^1_{b}\neq k^2_{b}=0$ or 1 if $k^2_{b}\neq k^1_{b}=0$. Similar to SNS-TFQKD, only the key bit when Alice and Bob decide to send a coherent state but happen to send out the single-photon state is considered genuinely valid coded bit. We give the equivalent entanglement-based scheme as follows:
\begin{align}
	\rho_{ABa_1a_2b_1b_2}
	&=\frac{\ket{0}_\text{A}\ket{01}_{a_1a_2}+\ket{1}_\text{A}\ket{10}_{a_1a_2}}{\sqrt{2}}
	\otimes
	\frac{\ket{0}_\text{B}\ket{10}_{b_1b_2}+\ket{1}_\text{B}\ket{01}_{b_1b_2}}{\sqrt{2}}
\end{align}
We can reformulate $\rho_{ABa_1a_2b_1b_2}$ in the $X_{\delta_a}=\{\ket{+\delta_a}_\text{A}=\frac{1}{\sqrt{2}}(\ket{0}_\text{A}+\e^{-\ii\delta_a}\ket{1}_\text{A}), \ket{-\delta_a}_\text{A}=\frac{1}{\sqrt{2}}(\ket{0}_\text{A}-\e^{-\ii\delta_a}\ket{1}_\text{A})\}$ and $X_{\delta_b}=\{\ket{+\delta_b}_\text{B}=\frac{1}{\sqrt{2}}(\ket{0}_\text{B}+\e^{-\ii\delta_b}\ket{1}_\text{B}), \ket{-\delta_b}_\text{B}=\frac{1}{\sqrt{2}}(\ket{0}_\text{B}-\e^{-\ii\delta_b}\ket{1}_\text{B})\}$ basis as
\begin{align}
	\rho_{ABa_1a_2b_1b_2}
	&=\frac{\frac{\ket{+\delta_a}_\text{A}+\ket{-\delta_a}_\text{A}}{\sqrt{2}}\ket{01}_{a_1a_2}+\frac{\ket{+\delta_a}_\text{A}-\ket{-\delta_a}_\text{A}}{\sqrt{2}\e^{-\ii\delta_a}}\ket{10}_{a_1a_2}}{\sqrt{2}} \nonumber\\
	&\quad\otimes
	\frac{\frac{\ket{+\delta_b}_\text{B}+\ket{-\delta_b}_\text{B}}{\sqrt{2}}\ket{10}_{b_1b_2}+\frac{\ket{+\delta_b}_\text{B}-\ket{-\delta_b}_\text{B}}{\sqrt{2}\e^{-\ii\delta_b}}\ket{01}_{b_1b_2}}{\sqrt{2}} \nonumber\\
	&=\frac{\ket{+\delta_a}_\text{A}\frac{\ket{01}_{a_1a_2}+\e^{\ii\delta_a}\ket{10}_{a_1a_2}}{\sqrt{2}}+\ket{-\delta_a}_\text{A}\frac{\ket{01}_{a_1a_2}-\e^{\ii\delta_a}\ket{10}_{a_1a_2}}{\sqrt{2}}}{\sqrt{2}}\nonumber\\
	&\quad\otimes
	\frac{\ket{+\delta_b}_\text{B}\frac{\ket{10}_{b_1b_2}+\e^{\ii\delta_b}\ket{01}_{b_1b_2}}{\sqrt{2}}+\ket{-\delta_b}_\text{B}\frac{\ket{10}_{b_1b_2}-\e^{\ii\delta_b}\ket{01}_{b_1b_2}}{\sqrt{2}}}{\sqrt{2}}.\label{rho_ABa1a2b1b2}
\end{align}
Note that we further consider the definition of phase error rate. After Charlie's measurement and announcement, Alice and Bob can also measure the ancillary bipartite qubits $\rho_\text{AB}$ in the $Z_\text{A}^\prime=Z_\text{B}^\prime=\{\ket{0},\e^{-\ii\delta_a}\ket{1}\}$ basis to generate key bit or measure $\rho_\text{AB}$ in the $X_{A}=X_{B}=\{\ket{+}=\frac{1}{\sqrt{2}}(\ket{0}+\e^{-\ii\delta_a}\ket{1}), \ket{-}=\frac{1}{\sqrt{2}}(\ket{0}-\e^{-\ii\delta_a}\ket{1})\}$ basis for phase error test. Similar to before, phase error rate can also be defined in the $X_{A}=\{\ket{+}=\frac{1}{\sqrt{2}}(\ket{0}+\e^{-\ii\delta_a}\ket{1}), \ket{-}=\frac{1}{\sqrt{2}}(\ket{0}-\e^{-\ii\delta_a}\ket{1})\}$ and $X_{\delta,B}=\{\ket{+\delta}=\frac{1}{\sqrt{2}}(\ket{0}+\e^{-\ii(\delta_a+\delta)}\ket{1}), \ket{-\delta}=\frac{1}{\sqrt{2}}(\ket{0}-\e^{-\ii(\delta_a+\delta)}\ket{1})\}$ basis and there is the same correlation between the two definitions as show in (\ref{e_delta}).
In order to facilitate understanding of the definition of phase error rate, here we imagine the following scenario according to (\ref{rho_ABa1a2b1b2}).

Alice generates a key bit $\kappa_a\in\{0,1\}$ and prepares the quantum state $\frac{\ket{01}_{a_1a_2}+\e^{\ii(\delta_a+\kappa_a\pi)}\ket{10}_{a_1a_2}}{\sqrt{2}}$. Bob also generates a key bit $\kappa_b\in\{0,1\}$ and prepares the quantum state $\frac{\ket{01}_{b_1b_2}+\e^{-\ii(\delta_b+\kappa_b\pi)}\ket{10}_{b_1b_2}}{\sqrt{2}}$. Note that $\delta_a\in[0,2\pi)$ and $\delta_b\in[0,2\pi)$ are predetermined. They both send the quantum states to Charlie for interference measurement to share key bit. According to complementarity \cite{RN127}, the phase error rate of genuinely valid coded bits in MP-QKD is the bit error rate in such a scenario. Note that achieving the prepared quantum states in the imagined scenario poses challenges. Consequently, Alice and Bob send phase-randomized coherent states with identical intensity, which allows them to estimate the phase error rate using the decoy-state method. The phase post-selection condition in the original MP-QKD protocol corresponds to the case that $|\delta_a-\delta_b|\le\Delta$ or $|\delta_a-\delta_b-\pi|\le\Delta$ here. We define $\delta=\delta_b-\delta_a$. Note that $\delta$ and $\delta+\pi$ are equivalent in phase error test. The original MP-QKD protocol provides an estimation of the following loose phase error rate
\begin{align}
	e_\text{ph}^\Delta
	&=\frac{1}{4\pi\Delta}\int_{0}^{2\pi}\int_{\delta_a-\Delta}^{\delta_a+\Delta}
	e_\text{ph}^{\delta_a,\delta_b} \dd\delta_b \dd\delta_a \nonumber\\
	&=\frac{1}{4\pi\Delta}\int_{0}^{2\pi}\int_{-\Delta}^{\Delta}
	e_\text{ph}^{\delta_a,\delta_a+\delta} \dd\delta \dd\delta_a ,\label{eph^Delta}
\end{align}
where $e_\text{ph}^{\delta_a,\delta_b}$ is the bit error rate in the imagined scenario for the given $\delta_a$ and $\delta_b$ as well as the phase error rate in the $X_{\delta_a}$ and $X_{\delta_b}$ basis. 

Similar to (\ref{e_delta}), we can also get the following equality
\begin{align}
	e_\text{ph}^{\delta_a,\delta_a+\delta}
	&=
	e_\text{ph}^{\delta_a,\delta_a}\cos\delta+\frac{1-\cos\delta}{2}+ (\sum_{j=0}^{s} p_{j}A_X^j)\sin\delta.\label{eph_delta_ab}
\end{align}
By integrating $\delta$ in the interval $[-\Delta,\Delta]$ and $\delta_a$ in the interval $[0,2\pi]$ in both sides of (\ref{eph_delta_ab}), we can obtain the following precise phase error rate
\begin{align}
	e_\text{ph}
	&=\frac{1}{2\pi}\int_{0}^{2\pi}
	e_\text{ph}^{\delta_a,\delta_a} \dd\delta_a \nonumber\\
	&= \frac{1}{\sinc\Delta}e_\text{ph}^{\Delta}+\frac{1}{2}\big(1-\frac{1}{\sinc\Delta}\big) \nonumber\\
	&\le \frac{1}{\sinc\Delta}\bar{e}_\text{ph}^{\Delta}+\frac{1}{2}\big(1-\frac{1}{\sinc\Delta}\big),
\end{align}
where $e_\text{ph}^{\Delta}$ is defined in (\ref{eph^Delta}) and $\bar{e}_\text{ph}^{\Delta}$ is the upper bound of $e_\text{ph}^{\Delta}$.

\begin{figure}
	\centering
	\includegraphics[width=\textwidth]{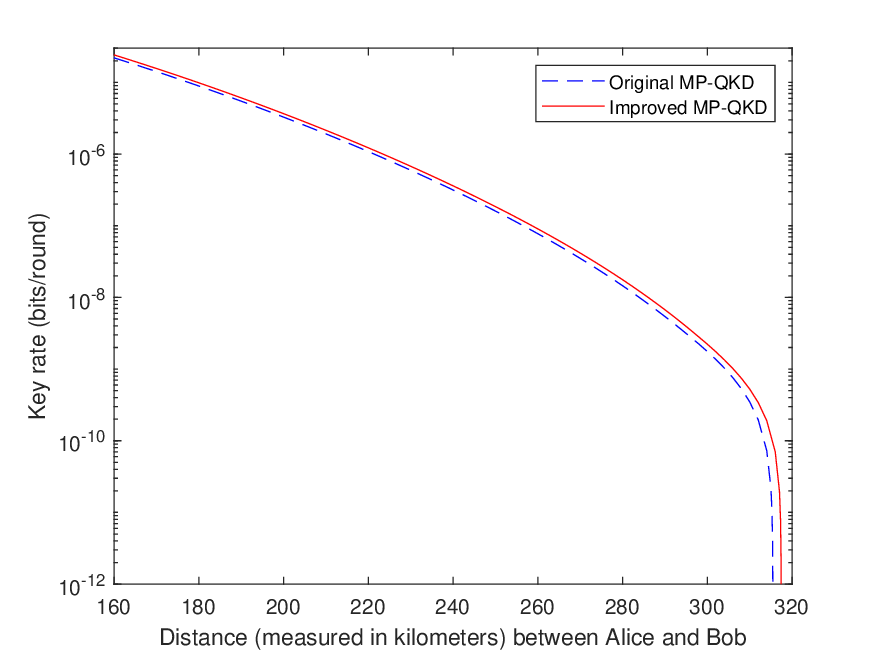}
	\caption{The key rate per round in MP-QKD with original phase error rate and our precise analysis.}
	\label{fig.MP_QKD}
\end{figure}

We use this precise phase error analysis in the practical MP-QKD protocol \cite{RN330} and simulate the original MP-QKD and our improved protocol in fig.\ref{fig.MP_QKD}.

We use the same finite-key analysis and simulation model mentioned in ref.\cite{RN330}. We set the total sending pulse as \num{1e13}, the maximal pairing interval as \num{1e2}, the misalignment error in Z basis as 0.5\%, the misalignment error in X basis as 5\%, the dark count rate as \num{1e-8}, the fiber loss coefficient as \SI{0.2}{dB/km}, the photon detection efficiency as 70\%, the error correction inefficiency as \num{1.1} and the failure probability when calculating the effect of statistical fluctuation as \num{1e-23}. Furthermore, we achieve a security level of \num{1e-10} in the sense of composable security against coherent attacks. The simulation results indicate a key rate improvement of over 10\% for distances beyond \SI{160}{km}, accompanied by an extended maximum achievable distance of \SI{2}{km}, which once again confirms that our precise phase error analysis noticeably improves the key rate performance in practice.

\section{CONCLUSION}
In conclusion, our precise phase error rate analysis provides a comprehensive and accurate comprehension of phase error rate for QKD with phase postselection. The versatility of our method enables its direct integration into AOPP-SNS-TFQKD and MP-QKD protocols, facilitating notable enhancements in key rate performance without necessitating alterations to the existing experimental hardware or protocol steps. Given its adaptable nature, we anticipate its applicability to extend beyond these specific protocols, offering potential improvements in various other QKD with phase postselection.

\appendix*
\section{A more intuitive phase error analysis for SNS-TFQKD}
In the equivalent entanglement-based scheme of SNS-TFQKD protocol,
we can reformulate $\sigma^\text{u}_\text{ABab}$ in (\ref{sigma^u_ABab}) under the $X_\delta=\{\ket{+\delta}=\frac{1}{\sqrt{2}}(\ket{00}+\e^{-\ii\delta}\ket{11}), \ket{-\delta}=\frac{1}{\sqrt{2}}(\ket{00}-\e^{-\ii\delta}\ket{11})\}$ basis ($0\le\delta<\frac{\pi}{2}$) as
\begin{align}
	\sigma^\text{u}_\text{ABab}
	&=\frac{\frac{\ket{+\delta}_\text{AB}+\ket{-\delta}_\text{AB}}{\sqrt{2}}\ket{01}_\text{ab}+\frac{\ket{+\delta}_\text{AB}-\ket{-\delta}_\text{AB}}{\sqrt{2}\e^{-\ii\delta}}\ket{10}_\text{ab}}{\sqrt{2}} \nonumber\\
	&=\frac{\ket{+\delta}_\text{AB}\frac{\ket{01}_\text{ab}+\e^{\ii\delta}\ket{10}_\text{ab}}{\sqrt{2}}+\ket{-\delta}_\text{AB}\frac{\ket{01}_\text{ab}-\e^{\ii\delta}\ket{10}_\text{ab}}{\sqrt{2}}}{\sqrt{2}}.
\end{align}
Here we follow the idea proposed in \cite{PhysRevA.98.062323} that the genuinely valid coded quantum state and the conjugated quantum state used for phase error estimation in SNS-TFQKD are similar to that in a BB84 protocol \cite{BB84}.
This indicates that the phase error rate defined in the $X_\delta$ basis is related to the yields of the quantum states $\frac{\ket{01}_\text{ab}+\e^{\ii\delta}\ket{10}_\text{ab}}{\sqrt{2}}$ and $\frac{\ket{01}_\text{ab}-\e^{\ii\delta}\ket{10}_\text{ab}}{\sqrt{2}}$. 

To estimate the phase error rate in the practical SNS-TFQKD protocol, Charlie should measure the bipartite signal qubits $\frac{\ket{01}_\text{ab}+\e^{\ii\theta}\ket{10}_\text{ab}}{\sqrt{2}}$ ($\theta\in[0,2\pi)$) sent from Alice and Bob on the measuring device $\mathcal{M}$. Since Charlie does not know the phase difference $\theta$ when measuring the signal qubits a and b, we can assume without loss of generality that Charlie's measurement device $\mathcal{M}$ can perfectly discriminate the quantum state between $\frac{\ket{01}_\text{ab}+\ket{10}_\text{ab}}{\sqrt{2}}$ and $\frac{\ket{01}_\text{ab}-\ket{10}_\text{ab}}{\sqrt{2}}$, i.e., the click of the L port must indicate the quantum state $\frac{\ket{01}_\text{ab}+\ket{10}_\text{ab}}{\sqrt{2}}$ and the click of the R port must indicate the quantum state $\frac{\ket{01}_\text{ab}-\ket{10}_\text{ab}}{\sqrt{2}}$. Naturally, Charlie can not perfectly discriminate whether the state is $\frac{\ket{01}_\text{ab}+\e^{\ii\delta}\ket{10}_\text{ab}}{\sqrt{2}}$ or $\frac{\ket{01}_\text{ab}-\e^{\ii\delta}\ket{10}_\text{ab}}{\sqrt{2}}$ ($\delta\neq 0$ or $\pi$), i.e., the click of the L port partially indicates the quantum state $\frac{\ket{01}_\text{ab}+\e^{\ii\delta}\ket{10}_\text{ab}}{\sqrt{2}}$ and the click of the R port partially indicates the quantum state $\frac{\ket{01}_\text{ab}-\e^{\ii\delta}\ket{10}_\text{ab}}{\sqrt{2}}$.
For the L port click events, we define the events caused by the quantum state $\frac{\ket{01}_\text{ab}-\e^{\ii\delta}\ket{10}_\text{ab}}{\sqrt{2}}$ as the error events, which enlightens us to define the quantum state $\ket{-\delta}_\text{AB}$ as an error in the $X_\delta$ basis.  Correspondingly, we define the events caused by the quantum state $\frac{\ket{01}_\text{ab}-\e^{\ii\delta}\ket{10}_\text{ab}}{\sqrt{2}}$ as the error events for the R port click events, which enlightens us to define the quantum state $\ket{+\delta}_\text{AB}$ as an error in the $X_\delta$ basis.

We first consider the L port click events announced by Charlie. Alice and Bob prepare the joint quantum state $\rho_\text{ABab}=\sigma_\text{ABab}^\text{u}$ in (\ref{sigma^u_ABab}) and send the signal qubits a and b to Charlie for measuring. We define the phase error rate $e_\text{ph}^{\delta,\text{L}}$ as
\begin{align}
	e_\text{ph}^{\delta,\text{L}}
	&=\bra{-\delta}\rho_\text{AB}\ket{-\delta} \nonumber\\
	&=\frac{1}{2}(\bra{00}-\e^{\ii\delta}\bra{11})\rho_\text{AB}(\ket{00}-\e^{-\ii\delta}\ket{11}) \nonumber\\
	&=\frac{1}{2}\big[
	\bra{00}\rho_\text{AB}\ket{00}+\bra{11}\rho_\text{AB}\ket{11}-\e^{-\ii\delta}\bra{00}\rho_\text{AB}\ket{11}-\e^{\ii\delta}\bra{11}\rho_\text{AB}\ket{00}
	\big] \nonumber\\
	&=\frac{1}{2}-\text{Re}[\e^{\ii\delta}\bra{11}\rho_\text{AB}\ket{00}],
\end{align}
where, \text{Re}[z] is the real part of the complex number $z$ and $\rho_\text{AB}$ is the quantum state of local auxiliary qubits A and B after Charlie's measurement on signal states a and b.

In the decoy mode of SNS-TFQKD, Alice and Bob send the coherent states to Charlie for interference on the beamsplitter, followed by two photon
detectors L and R. This allows us to estimate the yield of the quantum state $\rho_\Delta=\frac{1}{\Delta}\int_{-\frac{\Delta}{2}}^{\frac{\Delta}{2}}\big[ (\text{P}\{\frac{\ket{01}_\text{ab}+\e^{-\ii\delta}\ket{10}_\text{ab}}{\sqrt{2}}\}+\text{P}\{\frac{\ket{01}_\text{ab}-\e^{-\ii\delta}\ket{10}_\text{ab}}{\sqrt{2}}\})/2 \big]\dd\delta$ which indicates the phase error rate $e_\text{ph}^{\Delta,\text{L}}$ for the L port click events.
\begin{align}
	e_\text{ph}^{\Delta,\text{L}}
	&=\frac{1}{\Delta}\int_{-\frac{\Delta}{2}}^{\frac{\Delta}{2}}e_\text{ph}^{\delta,\text{L}} \dd\delta \nonumber\\
	&=\frac{1}{\Delta}\int_{-\frac{\Delta}{2}}^{\frac{\Delta}{2}} 
	\bigg(
	\frac{1}{2}-\text{Re}[\e^{\ii\delta}\bra{11}\rho_\text{AB}\ket{00}]
	\bigg) \dd\delta \nonumber\\
	&=\frac{1}{2}-\sinc(\frac{\Delta}{2})\text{Re}[\bra{11}\rho_\text{AB}\ket{00}],\label{eph-Delta-L}
\end{align}
where $\sinc(x)=\frac{\sin(x)}{x}$. In fact, $e_\text{ph}^{\Delta,\text{L}}$ is not the optimal phase error rate definition. We can use the current measurement data to estimate the precise phase error rate $e_\text{ph}^{\text{L}}$ in the $X_0=\{\ket{+}=\frac{1}{\sqrt{2}}(\ket{00}+\ket{11}), \ket{-}=\frac{1}{\sqrt{2}}(\ket{00}-\ket{11})\}$ basis as follows
\begin{align}
	e_\text{ph}^{\text{L}}
	&=\bra{-}\rho_\text{AB}\ket{-} \nonumber\\
	&=\frac{1}{2}(\bra{00}-\bra{11})\rho_\text{AB}(\ket{00}-\ket{11}) \nonumber\\
	&=\frac{1}{2}\big[
	\bra{00}\rho_\text{AB}\ket{00}+\bra{11}\rho_\text{AB}\ket{11}-\bra{00}\rho_\text{AB}\ket{11}-\bra{11}\rho_\text{AB}\ket{00}
	\big] \nonumber\\
	&=\frac{1}{2}-\text{Re}[\bra{11}\rho_\text{AB}\ket{00}].\label{eph-L}
\end{align}
Combining (\ref{eph-Delta-L}) and (\ref{eph-L}), we have
\begin{align}
	e_\text{ph}^{\text{L}}=\frac{1}{\sinc(\frac{\Delta}{2})}e_\text{ph}^{\Delta,\text{L}}+\frac{1}{2}\big(1-\frac{1}{\sinc(\frac{\Delta}{2})}\big).\label{ephL}
\end{align}

Similarly, we define the phase error rate $e_\text{ph}^{\delta,\text{R}}$ for the R port click events as
\begin{align}
	e_\text{ph}^{\delta,\text{R}}
	&=\bra{+\delta}\rho_\text{AB}\ket{+\delta} \nonumber\\
	&=\frac{1}{2}(\bra{00}+\e^{\ii\delta}\bra{11})\rho_\text{AB}(\ket{00}+\e^{-\ii\delta}\ket{11}) \nonumber\\
	&=\frac{1}{2}\big[
	\bra{00}\rho_\text{AB}\ket{00}+\bra{11}\rho_\text{AB}\ket{11}+\e^{-\ii\delta}\bra{00}\rho_\text{AB}\ket{11}+\e^{\ii\delta}\bra{11}\rho_\text{AB}\ket{00}
	\big] \nonumber\\
	&=\frac{1}{2}+\text{Re}[\e^{\ii\delta}\bra{11}\rho_\text{AB}\ket{00}].
\end{align}
Alice and Bob also use the sent quantum state $\rho_\Delta=\frac{1}{\Delta}\int_{-\frac{\Delta}{2}}^{\frac{\Delta}{2}}\big[ (\text{P}\{\frac{\ket{01}_\text{ab}+\e^{\ii\delta}\ket{10}_\text{ab}}{\sqrt{2}}\}+\text{P}\{\frac{\ket{01}_\text{ab}-\e^{\ii\delta}\ket{10}_\text{ab}}{\sqrt{2}}\})/2 \big]\dd\delta$ to estimate the phase error rate $e_\text{ph}^{\Delta,\text{R}}$ for the R port click events as follows
\begin{align}
	e_\text{ph}^{\Delta,\text{R}}
	&=\frac{1}{\Delta}\int_{-\frac{\Delta}{2}}^{\frac{\Delta}{2}}e_\text{ph}^{\delta,\text{R}} \dd\delta \nonumber\\
	&=\frac{1}{\Delta}\int_{-\frac{\Delta}{2}}^{\frac{\Delta}{2}} 
	\bigg(
	\frac{1}{2}+\text{Re}[\e^{\ii\delta}\bra{11}\rho_\text{AB}\ket{00}]
	\bigg) \dd\delta \nonumber\\
	&=\frac{1}{2}+\sinc(\frac{\Delta}{2})\text{Re}[\bra{11}\rho_\text{AB}\ket{00}].\label{eph-Delta-R}
\end{align}
The precise phase error rate $e_\text{ph}^{\text{R}}$ is
\begin{align}
	e_\text{ph}^{\text{R}}
	&=\bra{+}\rho_\text{AB}\ket{+} \nonumber\\
	&=\frac{1}{2}(\bra{00}+\bra{11})\rho_\text{AB}(\ket{00}+\ket{11}) \nonumber\\
	&=\frac{1}{2}\big[
	\bra{00}\rho_\text{AB}\ket{00}+\bra{11}\rho_\text{AB}\ket{11}+\bra{00}\rho_\text{AB}\ket{11}+\bra{11}\rho_\text{AB}\ket{00}
	\big] \nonumber\\
	&=\frac{1}{2}+\text{Re}[\bra{11}\rho_\text{AB}\ket{00}].\label{eph-R}
\end{align}
Combining (\ref{eph-Delta-R}) and (\ref{eph-R}), we also have
\begin{align}
	e_\text{ph}^{\text{R}}=\frac{1}{\sinc(\frac{\Delta}{2})}e_\text{ph}^{\Delta,\text{R}}+\frac{1}{2}\big(1-\frac{1}{\sinc(\frac{\Delta}{2})}\big). \label{ephR}
\end{align}

In the current SNS-TFQKD system, we usually consider the click events of the L and R ports together to calculate the total phase error rate
\begin{align}
	e_\text{ph}^\text{tot}=\frac{n_\text{L}}{n_\text{L}+n_\text{R}}e_\text{ph}(\text{L})+\frac{n_\text{R}}{n_\text{L}+n_\text{R}}e_\text{ph}(\text{R}),
\end{align}
where $n_\text{L}$ and $n_\text{R}$ are the number of click events from port L and R. The loose phase error rate $e_\text{ph}^\Delta=\frac{n_\text{L}}{n_\text{L}+n_\text{R}}e_\text{ph}^{\Delta,\text{L}}+\frac{n_\text{R}}{n_\text{L}+n_\text{R}}e_\text{ph}^{\Delta,\text{R}}$. The precise phase error rate $e_\text{ph}^\text{p}=\frac{n_\text{L}}{n_\text{L}+n_\text{R}}e_\text{ph}^{\text{L}}+\frac{n_\text{R}}{n_\text{L}+n_\text{R}}e_\text{ph}^{\text{R}}$. Combining (\ref{ephL}) and (\ref{ephR}), we can get the precise phase error rate from the loose phase error rate given the measurement data
\begin{align}
	e_\text{ph}^\text{p}=\frac{1}{\sinc(\frac{\Delta}{2})}e_\text{ph}^{\Delta}+\frac{1}{2}\big(1-\frac{1}{\sinc(\frac{\Delta}{2})}\big). \label{precise eph}
\end{align}


\begin{acknowledgments}
We acknowledge Dr. Rong Wang for the enlightening discussions.
This work has been supported by the National Key Research and Development Program of China (Grant No. 2020YFA0309802), the National Natural Science Foundation of China (Grant Nos. 62171424, 62271463), Prospect and Key Core Technology Projects of Jiangsu provincial key R \& D Program (BE2022071), the Fundamental Research Funds for the Central Universities, the Innovation Program for Quantum Science and Technology (Grant No. 2021ZD0300701).
\end{acknowledgments}

\bibliography{ref}

\end{document}